\documentclass [12pt] {article}
\usepackage {amssymb, amsfonts}
\newcommand{\sh}{\sinh}
\newcommand{\ch}{\cosh}
\begin {document}
\vskip0.3cm \centerline {{\bf { Exact Solutions and hypothesis on
Phase Transition}}} \centerline {\bf {in the Polyelectrolite Model
of DNA{\footnote {The work is executed at financial support of the
Russian Found of Basic Researches (project No
00-01-00480)}}}}\vskip0.4cm \hfil {\bf E.Sh.Gutshabash}\hfil
\vskip0.6cm \centerline {Institute Research for Physics,
Sankt-Petersburg State University,}\centerline {Russia, e-mail:
gutshab@EG2097.spb.edu} \vskip0.5cm \parbox {14cm} {\small The
two-dimensional generalization of the Polyelectolite model of DNA
is proposed. It is reduced to the boundary problem for nonlinear
quite integrable equation $\sh$-Gordon. In the linearisible
version the exact solution is constructed and its asymptotic is
found. The soliton solutions of nonlinear equation are calculated
that allowed tells about the possibility of the structural phase
transition in the considered system (DNA+polyelectrolite) on the
temperature.}

 \vskip0.6cm

At the present time constructing and investigating mathematical models
of DNA is an important problem of both theoretical biophysics and
mathematical physics. Modern achievements in the field of nonlinear
equations theory that occur, as a rule, in these models allow not only
to set initial and boundary problems for such equations, but also to
solve them in a number of cases.

The given work is devoted to construction of exact solutions for a
two-dimensional generalization of so-called polyelectrolyte model
of DNA (PM DNA) [1]. In essence here we deal with solution of an
inverse problem in the broad sense, i.e. it is to judge structure
and properties of a DNA molecule from a field restoring within the
framework of this model and also values of some functionals of it.

It is necessary to note that PM represents the most concentrated
description of role of Coulomb interactions which are of a great
importance in formation of structure of a DNA molecule and in its
functioning as well as in intrinsical structural transitions. One more
important advantage of the model is possibility to operate with terms
of statistical physics, which allows to implement a clear mathematical
problem statement.

{\bf The formulation of model and statement of the problem.}  In
reality a DNA molecule is a strongly charged poly-ion, with two
charges of electrons falling at every pair of nucleotides. The
idea of PM consists in replacement of such system by an infinite
and regular negatively charged cylinder of radius $r_0$ with a
given surface charge density placed into solution of
polyelectrolite. It is supposed that a processes of relaxation
proceeds quickly enough, so only an equilibrium situation is
considered. Besides here we are restricted ourselves by
consideration of the case when a solution consists of unitary
charged ions and electrons.

Standard physical reasons result to a system of Poisson-Boltsmann
equations written down at any point of volume outside the
cylinder:
$${\mathrm div}\: {\bf E} = 4\pi\rho, \:\: {\bf E} =
-{\mathrm grad}\: U, \:\:\rho = e (n _ {i}-n _ {e}), \eqno (1)
$$
in this case $ n _ {i, e}$ being under condition of thermodynamic
balance are defined by Boltsmann's distributions:

$$ n_i (U) = n _ {0i} e ^ {-\frac {eU} {T}}, \:\:\:n _ {e} (U) = n _
{0e} e ^ {\frac {eU} {T}} .\eqno (2) $$ In formulas (1), (2) the
following notations are used: $ {\bf E} = {\bf E} (x, y, z) $ is a
vector of electrical field intensity, $U = U (x, y, z)$ is a
dimensional scalar potential, $e$ is a charge of electron, $ \rho
$ is a charge density, $n_i, \:n_e$ are concentrations of ions and
electrons at the point ${\bf r} = \{x, y, z \}$ correspondingly,
$n_{0i}, \:n_{0e}$ are equilibrium concentrations, $T$ is absolute
temperature, Boltsmann's constant being assumed to be equal to
unit.

Supposing that $ n_{0i}=n_{0e} \equiv n_0$ (the condition of total
system neutrality) as well as that there is a symmetry in shifting
along an axis $z$ being a symmetry axis of the cylinder, we will
pass the plane orthogonal to this axis. Then introducing
dimensionless variables ${\bf r}^{\prime} = {\bf r}/ (2r _ D), \:u
= eU/T $, where $r_D$ is Debye radius, $r_D^2 = T / (8\pi en_0)$,
from (1), (2) we will obtain a basic equation of the model
(strokes at independent variables are omitted)

$$
\triangle u = 4\sinh u, \eqno (3)
$$
were $\triangle$ is two-dimensional Laplace's operator.

It should be noticed, that the equation (3) also can be obtained
from a strict and valid procedure of breakage and disengagement of
BBGKI chain in neglecting fluctuations [2,3], with the function $
u $ being an average potential of the system which is a functional
of Coulomb's two-partial potential. The equation (3), as applied
to PM DNA, corresponds to the case of a salt excess in a
polyelectrolyte solution {\footnote {If the right part of (3) to
substitute for an exponent (having a negative sign both before its
argument and before the exponent) will have Liouville's equation
having an exact solution [4] that corresponds to a salt lack).}}.

$ \sinh $ -Gordon equation (3) belongs to the class of
two-dimensional completely integrable equations of the elliptical
type, for which representation of zero curvature is known [5, 6]
and, hence, the Inverse Scattering Transform Method (IST) can be
applied. In the papers [7,8] this equation was solved for a
half-plane $y > 0$, in [9] an IST procedure was employed in
solving $\sinh$-Gordon-equation fitting Coulomb's system on a
plane at negative temperatures, and in a series of papers [10-12]
an equation of the same type (with a function $ \sin u$) was
solved on a whole plane with given conditions on chosen beams (for
the purpose of obtaining unambiguous solutions).

Further it will be convenient to work with the equation (3)
rewritten in polar coordina\-\-tes:

$$ u_{rr} + \frac {1} {r} u _ r + \frac {1} {r ^ 2} u _ {\varphi
\varphi} = 4\sinh u, \eqno (4) $$ where $ x = r \cos \varphi, \:y
= r \sin \varphi, \:r > r _ 0 $ (here $ r_0 $ is a dimensionless
radius of the cylinder) {\footnote {In the one-dimensional case
recently this equation has been numerically solved in [13].}}. Let
us assume that $ u = u (r, \varphi) $ is a real-valued function, $
u \to 0 $ with $ r \to \infty $ quickly enough, with $u(r,
\varphi) = u (r, \varphi + 2\pi) $.

 For (4) are natural two statements of boundary problems:

 a) Dirichlet's problem for an exterior of circle: $$ u_{| r = r _
0} = \hat f _ 1 (\varphi), \eqno (5) $$

b) Neumann's problem for an exterior of circle:
$$ { u _ r} _ {{| r =
r _ 0}} = \hat f _ 2 (\varphi), \eqno (6)
$$
in which $ \hat f _ 1 (\varphi), \:\hat f _ 2 (\varphi) $ are
given periodic functions (linear densities of charges of the
cylinder revealing a specific character of DNA molecule).

{ \bf Linearized model (a two-dimensional generalization of
Debay-Chukkel's model).} At $ r >> r_0 $ the equation (4) can be
linearized and then can be written as

$$
u_{rr} + \frac {1} {r} u _ r + \frac {1} {r ^ 2} u _
{\varphi\varphi} = 4u.
$$
Taking $ u = u (r, \varphi) = R (r) \Phi (\varphi) $, we shall
have

$$
r^2R _ {rr} + rR _ r - (4r ^ 2 + \lambda) R = 0, \:\:\:\: \Phi _
{\varphi\varphi} + \lambda\Phi = 0, \eqno (7)
$$
where a parameter $ \lambda\in \mathbb {C} $. Considering a
finiteness of solution from (7) we find:

$$ u(r, \varphi) = \sum _ m b _ mK _ m (2r) e ^ {im\varphi},
\:\:\:\lambda = m ^ 2. \eqno (8) $$ where $ b _ m $ are constants, $ K
_ m (.) $ are modified Bessel functions (McDonald's functions), and
the summation is over $ m $ within limits from $ -\infty $ to $ +
\infty $. In the case of the problem (5) coefficients $ b _ m $ are
defined from the condition

$$
b_m = \frac {1} {2\pi K _ m (2r _ 0)} \int _ 0 ^ {2\pi}
d\varphi\hat f _ 1 (\varphi) e ^ {-im\varphi}.
$$
Thus, we obtain

$$ u(r, \varphi) = \frac {1} {2\pi} \int _ 0 ^ {2\pi} d\varphi ^
{\prime} \sum _ m e ^ {im (\varphi-\varphi ^ {\prime})} \frac {K _
m (2r)} {K _ m (2r _ 0)} \hat f_1 (\varphi ^ {\prime}) .\eqno (9)
$$

Similarly in the case of a problem (6) we find $$ u(r, \varphi) =
\sum _ m c _ mK _ m (2r) e ^ {im\varphi}, \:\:\:\lambda = m ^ 2,
$$ and the coefficients $ c _ m $ are defined by relations

$$ c _ m = -\frac {1} {2\pi} \frac {\int _ 0 ^ {2\pi} d\varphi \hat
f_ 2 (\varphi) e ^ {-im\varphi}} {K _ {m-1} (2r _ 0) + K _ {m + 1}
(2r _ 0)}. $$

Hence it follows that

$$ u (r, \varphi) = -\frac {1} {2\pi} \int _ 0 ^ {2\pi} d\varphi ^
{\prime} \sum _ m e ^ {im (\varphi-\varphi ^ {\prime})} \frac { K
_ m (2r)} {K _ {m-1} (2r _ 0) + K _ {m + 1} (2r _ 0)} \hat f_2
(\varphi ^ {\prime}) .\eqno (10) $$

It is easy to show that both (9) and (10) meets the requirement of
reality. Representa\-\-tions (9) and (10) are solutions of
Debay-Chukkel's two-dimensional theory corresponding to
anisotropic medium (polyelectrolyte), which can arise owing to,
for example, possible (probable) fluctuations in the system.

Fourier-factors $ \hat f _ {1m}, \hat f _ {2m} $ of expansion of
functions $ \hat f _ 1 (\varphi), \hat f _ 2 (\varphi) $, are
obviously connected by the relations: $ \hat f _ {1m} /K _ m (2r _
0) = \hat f_{2m} / (2K _ m ^ {\prime} (2r _ 0)) $, which allows to
establish a nonlocal relation {\footnote {The idea of the approach
belongs to V.D.Lipovsky.}):

$$
u_ r (r _ 0, \varphi) = \int _ 0 ^ {2\pi} d\varphi ^ {\prime} H (r
_ 0, \varphi-\varphi ^ {\prime}) u (r _ 0, \varphi ^ {\prime}),
\eqno (11)
$$
in which the function $ H (r _ 0, \varphi) $ is determined by the
equality:

$$
H (r _ 0, \varphi) = \frac {1} {\pi} \sum _ m \frac { K _ m ^
{\prime} (2r)} {K _ m (2r _ 0)} e ^ {im\varphi}.
$$
Using (9) and the Fourier-factors connection (11) established above
we obtain

$$
u(r, \varphi) = \frac {1} {4\pi} \int _ 0 ^ {2\pi} d\varphi ^
{\prime} \sum _ m e^{im (\varphi- \varphi ^ {\prime})} K _ m (2r)
\Bigl [\frac {\hat f _ 1 (\varphi ^ {\prime})} {K _ m (2r _ 0)} +
\frac {\hat f _ 2 (\varphi ^ {\prime})} {2K _ m ^ {\prime} (2r _
0)} \Bigr].
$$
Returning to (8), where

$$
b _ m = \frac {1} {2\pi} \int _ 0 ^ {2\pi} d\varphi e ^
{-im\varphi} \Bigl [ \frac {u (r _ 0, \varphi)} { K _ m (2r _ 0)}
+ \frac {u _ r (r _ 0, \varphi)} {2K _ m ^ {\prime} (2r _ 0)}
\Bigl],
$$
and considering that $ b _ {-m} = {\bar b} _ m $, we have

$$
u (r, \varphi) = u _ 0 (r) + 4\sum _ {m = 1} ^ {\infty} ({\mathrm
Re} \; b _ m) K _ m (2r) \cos m\varphi, \eqno (12)
$$
with $u _ 0 (r) = b _ 0K _ 0 (2r), \:b _ 0 = {\bar b} _ 0 $. The
formula (12) is convenient for obtaining asymptotic
representations. Then

$$
u_ 0 (r) \sim \frac {b _ 0} {\sqrt {\pi r}} e ^ {-2r}, \:\: r \to
\infty.
$$
Thus, we can deduce a relation describing an effect of Debye
screening. Putting

$$
u ^{+ }(r, \varphi) = \sum _ {m = 1} ^ {\infty} {\tilde b} _ m K _
m (2r) e ^ {im\varphi}, \:\:\: {\tilde b} _ m = 4 {\mathrm Re} \;
b _ m,
$$
and assuming the possibility of analytic continuation of $ {\tilde
b}_m $ on an index $ m $, we convert a last sum in integral in the
ordinary way:

$$
u^{+} (r, \varphi) = \int _ C ds\: \frac {{\tilde b} _ se ^ {i
(\pi + \varphi) s}} {2i\sin \pi s} K_s (2r), \eqno (13)
$$
where a contour of integration starts from an infinite (along $
{\mathrm Re}\: s $) point located in the first quarter of the
complex variable $ s $ plane, then goes parallel to the real axis
crossing it at the point $ 0 < {\mathrm Re} \; s < 1 $, and then
ends at infinity parallel to the real axis and remaining in the
fourth quarter.

To calculation an asymptotic form of the expression (13) it should be
applied the saddle-point method. For that at first we should notice
that the function $ K _ s $ satisfies the equation

$$ K_{s, rr} + \frac {1} {r} K_{s, r} - (4 + \frac {m ^ 2} {r ^
2}) K _ s = 0. \eqno (14) $$ Taking $ \xi = 2r, \:K _ s = f _ s/\sqrt
{\xi} $, we reduce (14) to the form

$$
f _ s ^ {\prime \prime} - (\frac {m ^ 2-\frac {1} {4}} {\xi ^ 2} +
1) f _ s = 0.
$$
From here it follows that

$$ K_s(\xi) \sim \sqrt{\frac {2}{\pi \xi}} \Bigl(\frac
{s^2-\frac{1}{4}}{\xi^2}+1\Bigr)^{-\frac{1}{4}} \exp
\left\{-\sqrt{1+\frac{\xi^2}{s^2-\frac{1}{4}}}+\frac{1}{2}\ln
\frac{\sqrt{1+\frac{\xi^2}{s^2-\frac{1}{4}}}-1}{\sqrt{1+\frac{\xi^2}
{s^2-\frac{1}{4}}}+1}\right\},$$ or, after simple calculations
(assuming $ s/\xi = o (1), \:\xi \to \infty $), $$ K_s (2r) \sim
\frac {1} {\sqrt {\pi r}} e ^{-2r}e^{-\frac {s^2} {4r}},
\:\:\:\:\:\:\: r \to \infty. $$

Then it is possible to rewrite (13) as:

$$
u ^{+}(r, \varphi) \sim {\frac {1} {\sqrt {\pi r}}} e ^ {-2r} \int
_ C ds\frac {{\tilde b} _ s} {2i\sin \pi s} e ^ {is (\pi +
\varphi) -\frac {s ^ 2} {4r}}, \:\:r \to \infty .\eqno (15)
$$

Let us assume that $ s = s _ 0\xi, \: s _ 0 = 2i (\pi + \varphi) r $.
Then a whole picture will be turned through a corner - $ \pi/2 $: the
poles will turned out in the negative part of the imaginary axis, and
the integration contour $ C $ will turn into a contour $ C ^ {\prime}
$ beginning at the point $ a-i\infty $, going parallel to the
imaginary axis and then crossing it and finally ending at the point $
-a-i\infty $  remaining parallel of the axis, where $ a $ is a
positive constant. Assuming that "factors" \enskip $ {\tilde b} _ s  $
permit an analytic continuation to required areas of the lower
semiplane, instead of (15) we obtain

$$u ^{+}(r, \varphi) \sim \frac {1} {\sqrt {\pi r}} e ^ {-2r} s _
0\int _ {C ^ {\prime}} d\xi\frac {{\tilde b} (s _ 0\xi)} {2i\sin
[2i\pi ( \pi + \varphi) r\xi]} e ^ {-r (\pi + \varphi) ^ 2
(2\xi-\xi ^ 2)}, \:\: r \to \infty. $$

On introducing a function $ G (\xi) = \xi^2-2\xi $ we find the
saddle point: $ \xi _ 0 = 1 $; then in its vicinity $ G (\xi) = -1
+ (\xi-\xi _ 0) ^ 2 + O ((\xi-\xi) ^ 3) $. Deforming the contour
in such a way that it goes through the saddle point in the
direction of the steepest descent, we have

$$ u^{+}(r,\varphi)=\left \{\frac{{\tilde b}(2i(\pi+\varphi)r)}
{\sh[2r\pi(\pi+\varphi)]}(1+O(\frac{1}{r}))\right\}e^{-[2+(\pi+\varphi)^2]
r},
\:\:\:r \to \infty. $$

This expression, which is the leading member of the asymptotic
form, gives an amend\-\-ment to the effect of Debye screening on
plane.

{\bf Exact solutions and "soliton"\enskip configurations of the
equation (4).} Let us proceed to constructing exact solutions of
the equation (4). It should be notice that the solution of the
boundary problems (4), (5) or (4), (6) encounters serious
mathematical difficulties involving application of IST (a
scattering problem for the operator of an associated linear
problem on half-axis). Therefore here we should be restricted by a
more simple problem: in neglecting the radius $ r _ 0 $ we will
construct exact solutions of (4) on a whole plane by means of
Darboux Transformation (DT) method.

Employing a direct calculation it is possible to check up that the
equation (4) is a condition of compatibility of the overdetermined
linear matrix system

$$
\Psi _ r = U\Psi, \:\:\:\Psi _ {\varphi} = rV\Psi, \eqno (16)
$$
where $ \Psi = \Psi (r, \varphi, \lambda), \:U = U (r, \varphi,
\lambda), \:V = V (r, \varphi, \lambda) \in Mat (2, \mathbb {C}),
\lambda \in \mathbb {C} $ is a parameter, and the matrixes $ U,
\:V $ are given as {\footnote {For the first
time these expressions were obtained by S.S. Nikulichev}}:

$$
U (r, \varphi, \lambda) = \frac {i\lambda} {2} e ^ {i\varphi}
\sigma _ 3 + \frac {\cosh u} {2i\lambda} e ^ {-i\varphi} \sigma _
3-\frac {u _ r-\frac {i} {r} u _ {\varphi}} {4} \sigma _ 2-\frac
{\sinh u} {2\lambda} e ^ {-i\varphi} \sigma _ 1,
$$
$$
V (r, \varphi, \lambda) = -\frac {\lambda} {2} e ^ {i\varphi}
\sigma _ 3-\frac {\cosh u} {2\lambda} e ^ {-i\varphi} \sigma _
3-i\frac {u _ r-\frac {i} {r} u _ {\varphi}} {4} \sigma _ 2-\frac
{\sinh u} {2i\lambda} e ^ {-i\varphi} \sigma _ 1.
$$
Here $ \sigma _ i, \:i = 1,2,3, $ are Pauli's standard matrixes.

Further instead of variable $ r, \:\varphi $ it is convenient to
introduce the "conic"\enskip variables $ \zeta = (\varphi + i\ln
r) /2, \: {\bar \zeta} = (\varphi-i\ln r) /2 $, and also using
invariance of a condition of compatibility of the system (16)
relative to a cyclic rearrangement of Pauli's matrixes, it is
worth changing to the another its gauge. Then instead of (16) we
have the following $ 2\times 2 $ matric system

$$
\Psi _ {\zeta} = A\Psi, \:\:\:\Psi _ {\bar {\zeta}} = B\Psi, \eqno
(17)
$$
where

$$
A = A (\zeta, {\bar {\zeta}}, \lambda) = \frac {1} {\lambda} e ^
{-2i\zeta}
\left (\begin {array} {cc} 0 & e ^ u \\
                           e ^ {-u}& 0
\end {array} \right), \:\:\:
B=B(\zeta,{\bar {\zeta}},\lambda)= \left( \begin{array}{cc}
\frac{u_{\bar {\zeta}}}{2}&\lambda
e^{2i{\bar \zeta}}\\
\lambda e^{2i{\bar \zeta}}&-\frac{u_{\bar {\zeta}}}{2}
\end{array}\right).
$$
The compatibility condition (17) becomes $ A _ {\zeta} -B
_ {\bar {\zeta}} + [A, B] = 0, $ and an appropriate nonlinear equation,
takes the form

$$
u_{\zeta {\bar {\zeta}}} = 4\sinh u\: e ^ {-2i (\zeta - {\bar
{\zeta}})},
$$
which, as it is easy to show, is equivalent to (4).

Let $ \Psi = (\Psi ^ {(1)}, \Psi ^ {(2)}), \Psi ^ {(1)} = (\theta,
\chi)^T $. Then the equations system (17) can be written as a
system of four equations

$$
\theta _ {\zeta} = \frac {e ^ u} {\lambda} e ^ {-2i\zeta} \chi,
\:\:\:\: \chi _ {\zeta} = \frac {e ^ {-u}} {\lambda} e ^
{-2i\zeta} \theta,
$$
$$
\eqno (18)
$$
$$
\theta _ {{\bar \zeta}} = \frac {u _ {{\bar \zeta}}} {2} \theta +
\lambda e^{2i {\bar {\zeta}}} \chi, \:\:\:\: \chi _ {\bar {\zeta}}
= \lambda e^{2i {\bar \zeta}} \theta-\frac {u _ {\bar {\zeta}}}
{2} \chi.
$$
Let $ \theta _ 1, \:\chi _ 1 $ be a fixed solution of (18)
corresponding to the choice $ \lambda = \lambda _ 1 $. Let us assume

$$
\tilde \theta = \lambda \chi-\lambda _ 1\frac {\chi _ 1} {\theta _
1} \theta, \:\:\:\:\: \tilde \chi = \lambda \theta-\lambda _
1\frac {\theta _ 1} {\chi _ 1} \chi,
$$
and check up covariance of the system (18) relative to
DT of a such form. After simple calculations it can be obtained

$$
\tilde u = u + 2\ln {\frac {\chi _ 1} {\theta _ 1}} .\eqno (19)
$$
In the theory of DT the relation is called a "dressing"\enskip
one.

To construction an explicit solution it is necessary to assign
some initial one; here as the solution we choose $ u = 0 $. Turning
back to polar coordinates, then from system (18) we obtain

$$
\theta _ {1r} = -\frac {1} {2i} [\cos \varphi ( \frac {1} {\lambda
_ 1} -\lambda _ 1) -i\sin \varphi (\frac {1} {\lambda _ 1} +
\lambda _ 1)] \chi _ 1,
$$
$$
\eqno (20)
$$
$$
\theta _ {1\varphi} = \frac {1} {2} r [\cos \varphi ( \frac {1}
{\lambda _ 1} + \lambda _ 1) -i\sin \varphi (\frac {1} {\lambda _
1} -\lambda _ 1)] \chi _ 1.
$$
The given system has an integral: $ \theta _ 1 ^ 2-\chi _ 1 ^ 2 =
A_1^2 $, where $A_1$ is an arbitrary, generally speaking, complex
constant, which allows to integrate (20). We have

$$
\chi _ 1 = \frac {1} {2} A _ 1 (e ^ {\Gamma _ 1-\ln A _ 1} -e ^
{-\Gamma _ 1 + \ln A _ 1}), \:\: \theta _ 1 = \frac {1} {2} A _ 1
(e ^ {\Gamma _ 1-\ln A _ 1} + e ^ {-\Gamma _ 1 + \ln A _ 1}).
$$
Here $ \Gamma _ 1 = -1 / (2i) [\cos \varphi (1/\lambda _ 1-\lambda
_ 1) -i\sin \varphi (1/\lambda _ 1 + \lambda _ 1)] r + \ln B _ 1,
\:B _ 1 $ is a real constant ( for simplicity $ A _ 1 $  is
considered to be real as well).

Considering (19), it is can be seen that the potential becomes real at
$ \lambda _ 1 = e ^ {i\alpha _ 1}, \:\alpha _ 1\in \mathbb {R}, \:
\alpha _ 1 \in [0, 2\pi] $. Then the simplest solution of the
equations (4) takes the form ($ u [1] \equiv \tilde u $)

$$
u [1] = 2\ln \left \{\tanh [r\sin (\varphi + \alpha _ 1) + \delta
_ 1] \right \},\eqno (21)
$$
where $\delta _ 1 = \ln |B_1/A_1|$. The solution (21) is an
analogue of the solution type of 1-kink for nonlinear equations of
the hyperbolic type [5] and describes a distribution of the
potential on plane. Also it is of interest to note that in solving
the equation (3) on half-plane by means of IST, 1-kink solutions
turn out to be forbidden because of the potential must be real
[8].

Let us consider an double dressing procedure for the initial solution.
Repeating procedure of obtainment (19) we have

$$
u[2] = u + 2\ln \frac {\chi _ 1\chi _ 2 [1]} {\theta _ 1\theta _ 2
[1]} .\eqno (22)
$$
Here

$$ \theta_2 [1] = e ^ {-2i {\bar \zeta}} (\theta _ {2 {\bar \zeta}}
-\frac { \theta _ {1 {\bar \zeta}}} {\theta _ 1} \theta _ 2) = \lambda
_ 2\chi _ 2-\lambda _ 1\frac {\chi _ 1} { \theta _ 1} \theta _ 2, $$
$$ \chi _ 2 [1] = e ^ {-2i {\bar \zeta}} (\chi _ {2 {\bar \zeta}}
-\frac { \chi _ {1 {\bar \zeta}}} {\chi _ 1} \chi _ 2) = \lambda _
2\theta _ 2-\lambda _ 1\frac {\theta _ 1} { \chi _ 1} \chi _ 2, $$
and $ \chi _ 2 = \chi _ 2 (r, \varphi), \:\theta _ 2 = \theta _ 2
(r, \varphi) $ are solutions of the system (20) for $ \lambda =
\lambda _ 2$. Then the relation (22) can be rewritten in the form

$$ u [2] = u + 2\ln \frac { \lambda _ 2\chi _ 1\theta _ 2-\lambda _
1\theta _ 1\chi _ 2} { \lambda _ 2\chi _ 2\theta _ 1-\lambda _ 1\chi _
1\theta _ 2} .\eqno (23) $$

On examining realty of the expression (23) it can be resulted in two
possible cases.

1. Let $ \lambda_k =e^{i\alpha_k}, \:\alpha _ k \in [0,2\pi], \:k
= 1,2 $, i.e. both complex parameters must belong to an unit
circle {\footnote {Employing IST for half-plane it has been shown
that a parameter $ \lambda $ has sense of a spectral parameter for
an associated linear problem (16) [7,8] whereas values $ \lambda _
k $ form its discrete spectrum.}}. Let us assume $ \Gamma _ k =
r\sin (\varphi + \alpha _ k) + \ln B _ k, \:B _ k \in \mathbb {R},
\:k = 1, \:2 $, and as solutions of the system (34) take

$$ \chi _ k = C _ ke ^ {\Gamma _ k + \ln \delta _ k} + iD _ ke ^
{-\Gamma _ k-\ln \delta _ k}, \:\:\: \theta _ k = C _ ke ^ {\Gamma _ k
+ \ln \delta _ k} -iD _ ke ^ {-\Gamma _ k-\ln \delta _ k}. $$

In this case the realty is guaranteed provided $ \chi _ k = {\bar
\theta _ k}, \:C _ k, \:D _ k \in \mathbb {R}, \:-4iC _ kD _ k = A
_ k ^2$, and then we obtain:

$$ u_{ks} [2] = 2\ln \frac {1 + h _ {ks}}
{1-h _ {ks}}, \:\:h _ {ks} = h (r, \varphi) = -\cot Q\frac {\sinh
(\Gamma _ 2-\Gamma _ 1 + \nu _ 1)} {\cosh (\Gamma _ 1 + \Gamma _ 2 +
\nu _ 2)}. \eqno (24) $$ Here $ \nu _ 1 = \ln | \delta _ 2/\delta _ 1
| - (1/2) \ln | R _ 1/R _ 2 |, \:\:\nu _ 2 = \ln | \delta _ 1\delta _
2 | + (1/2) \ln | R _ 1R _ 2 |, \:\:\delta _ 1, \:\:\delta _ 2 \in
\mathbb {R}, \:\:R _ k = D _ k/C _ k, \:\:q = (\alpha _ 1-\alpha _ 2)
/2 $.

The solution (24) is an analogue of 2-kinks solution for equations of
the hyperbolic type. It differs a little from a solution of the
equation (3) obtained in [8] for half-plane by means of IST (ibidem
the appropriate asymptotic forms at $ r \to \infty $ have been found).

2. Let $ \lambda _ 1 = \gamma _ 1e ^ {i\alpha _ 1}, \:\lambda _ 2 = 1
/ {\bar \lambda _ 1}, \:\alpha _ 1, \:\gamma _ 1 \in \mathbb {R},
\:\alpha _ 1\in [0,2\pi], \:\gamma _ 1> 0 $. Then $ \Gamma (\lambda _
1) = {\bar \Gamma} (\lambda _ 2), \: \Gamma (\lambda _ 1) \equiv
\Gamma _ 1 = -1 / (2i) [\cos \varphi (1/\lambda _ 1-\lambda _ 1)
-i\sin \varphi (1/\lambda _ 1 + \lambda _ 1)] r = \Gamma _ {1R} +
i\Gamma _ {1I}, \: \Gamma _ {1R} = (1/2) \sin (\varphi + \alpha _ 1)
(1/\gamma _ 1 + \gamma _ 1) r, \: \Gamma _ {1I} = (1/2) \cos (\varphi
+ \alpha _ 1) (\gamma _ 1-1/\gamma _ 1) r $. Setting $ \theta _ 1 =
{\bar \chi _ 2}, \:\theta _ 2 = {\bar \chi _ 1}, \:k = 1,2 $, where $
\chi _ k = F _ ke ^ {\Gamma _ k} -G _ ke ^ {-\Gamma _ k}, \:\:\theta _
k = F _ ke ^ {\Gamma _ k} + G _ k E ^ {-\Gamma _ k}, $ we have the
other real solution:

$$
u_b [2] = 2\ln \frac {1-h _ b} {1 + h _ b}, \:\:\:h _ b = h (r,
\varphi) = \cot (\ln \gamma _ 1) \frac {\sin [2\cos (\varphi +
\alpha _ 1) (\sinh \gamma _ 1) r]} {\cosh [2\sin (\varphi + \alpha
_ 1) (\cosh \gamma _ 1) r + \mu _ 1]} .\eqno (25)
$$
Here $ F _ k, \:G _ k \in \mathbb {R}, \:4F _ kG _ k = A _ k ^ 2,
\:A _ k ^ 2 \in \mathbb {R}, \:\mu _ 1 = (1/2) \ln [(F _ 1 ^ 2 + F
_ 2 ^ 2) / (G _ 1 ^ 2 + G _ 2 ^ 2)] $.

The formula (25) is an analogue of the solution type of breather (a
double soliton) for a hyperbolic equation. In [8] the solution has
been obtained by help of IST and its asymptotic forms have been
calculated.

Procedure of obtaining subsequent "dressing"\enskip relations is
same as in deducing (23) (see, for example, [14]), and also
results in $ N $ - "soliton" \enskip solutions.

 {\bf On Phase Transition in PM DNA.} In the context of the problem
 under examination a rather wide solutions spectrum calculated above
 can be interpreted as one answering to various parameters or,
 perhaps, to various states of DNA. It allows, in particular, to put
 forward a hypothesis about a probable (within the framework of a used
 model) structural phase of transition of the 2-nd sort on temperature
 (for example, type of spiral $ \to $ ball or B-form $ \to $ Z-form,
 where B and Z are right and left spiral DNA accordingly). Such a
 transition is possible due to our system is two-dimensional one (it
 should be reminded that, according to Landau [15], such changes are
 forbidden in one-dimensional systems). As its mechanism the following
 one can be considered.

  Let two complex parameters (an eigen values) forming an inversion
  relative to an unit circle in the plane of a complex parameter $
  \lambda $ (state of the "breather"\enskip type) are available. It is
  assumed that in the system some change of temperature occurs, so
  that given parameters become dependent on it. It means that they
  start moving along the beam drawing between them (on the complex
  plane). In bringing together in the vicinity of the unit circle "a
  scattering" \enskip of the parameters occurs one on the other, and
  as a result they either with some probability $ p = p (\alpha, T) $
  drift apart along the circle, or with the probability $1-p$  "pass"
  \enskip through each other in such a way as to keep forming
  inversion relative to the circle obeying requirement of realty of
  the potential. The temperature, at which the phenomenon take places,
  is critical for the process; and the process corresponds to a
  structural phase of transition of the 2-nd sort {\footnote {In other
  words, here the phase of transition is connected to an asymptotical
  degeneration of eigen values, and the potential plays a part of the
  ordering parameter.}}. In order to prove this, following the idea of
  Landau, we are to calculate free energies of 2-kinks and breather
  configurations in the vicinity of the critical temperature and
  obtain a (nonzero) jump of the second derivative with respect to a
  free energy on temperature.

However, there is one essential difficulty. The point is that
according to so-called Derrick's theorem, in two-dimension systems
a functional of energy turns out to be infinite (see [16], where
questions on stability of integrable equations solutions have been
detailed considered as well). Therefore to estimate the free
energy functional, strictly speaking, it is to employ some
mathematically correct procedure. Nevertheless, we suggest quite a
simple (and correct, in our opinion) technique allowing to avoid
this difficulty.

Let us identify a free energy with a generating functional of the
model (4):

$$ {\cal F}(u)=\int \int_{\Bigl \{{\mathbb {R}^2 \setminus
C_{r=r_0}}\Bigr \}} dxdy\left\{4(\cosh u-1))+
\frac{1}{2}(u_x^2+u_y^2)\right\}=
$$
$$
=\int_{r=r_0}^{\infty}\int_0^{2\pi}dr d\varphi\:r\left\{ 4(\cosh
u-1 )+\frac{1}{2}(u_r^2+\frac{1}{r^2}u_{\varphi}^2) \right \},
$$
where a symbol $ {\Bigl \{\mathbb {R}^2 \setminus {C _ {r = r _
0}}}\Bigr \} $ means that one integrates over the space $ \mathbb
{R} ^ 2 $ with a remote circle of radius $ r _ 0 $. Using the
representations (24) or (25) for a function $ u = u (r, \varphi)
$, we  reduce $ {\cal F} (u) $ to the form convenient for
subsequent estimations:

$$ { \cal F} (u) = 8\int _ {r = r _ 0} ^ {\infty} \int _ 0 ^ {2\pi}
drd\varphi \: r\frac {4h ^ 2 + h _ r ^ 2 + \frac {h _ {\varphi} ^
2} {r ^ 2}} {(1 + h) ^ 2 (1-h) ^ 2}. \eqno (26) $$

Here an auxiliary function $ h = h (r, \varphi) $ satisfies a
nonlinear (and also, as well as (4)) a completely integrable equation

$$ h _ {rr} + \frac {1} {r ^ 2} h _ {\varphi\varphi} + \frac {2 (h _ r
^ 2 + \frac {h _ {\varphi} ^ 2} { r ^ 2})} {1-h ^ 2} + \frac {h _
r} {r} = \frac {4h (1 + h ^ 2)} {1-h ^ 2}, $$ whose representation
of a zero curvature can be obtained from the conditions of system
compatibility (16).

According to the critical phenomena theory  [17,18], close to the
critical point it is to assume $ \gamma _ 1 = \gamma _ 1 (t) = e ^
{v _ 1} $, where $v _ 1 =v_1 (t) \sim t ^ {d _ 1}, \: t = (T _
{\ast} -T) /T _ {\ast}, \: d _ 1 $ is the critical factor, $ d _
1> 0, \: T _ {\ast} $ is the critical temperature. Then a simple
estimation for $ h = h _ b $  near $ T = T _ {\ast} $ gives

 $$h_b \sim  \frac {1} {v _ 1} \frac {\sin [2\cos (\varphi +
\alpha _ 1) {\hat C} _ 1r]}  {\ch [2\cos (\varphi +  \alpha _ 1) {\hat
C} _ 2r + \mu _ 1]} \equiv \frac {1} {v _ 1} F _ b (r, \varphi), \:\:v
_ 1  \to 0.  $$

Here $ {\hat C} _ 1, \: {\hat C} _ 2 $ is a real constant.

Operating analogously for a condition characterizing by two
"kinks", on setting $v_2\equiv ( \alpha _ 1-\alpha _ 2) /2\sim t ^
{d _ 2} $, where $ d _ 2 $  a critical factor also, $ d _ 2> 0 $,
we find ( $ \Phi = \varphi + (\alpha _ 1 + \alpha _ 2) /2 $)

$$
h_{ks} \sim -\frac {1} {v _ 2} \frac {\sh \nu _ 1} {[\ch \nu _ 2 +
\sh (2r\sin \Phi) \sh \nu _ 2]} \equiv \frac {1} {v _ 2} F _ {ks}
(r, \varphi), \:\:v _ 2 \to 0.
$$
Thus, from (26) one can obtain

$$ { \cal F} _ b (t, r _ 0) \sim v _ 1 ^ 2Q _ b (r _ 0), \:\:\: {\cal
F} _ {ks} (t, r _ 0) \sim v _ 2 ^ 2Q _ {ks} (r _ 0), $$ where $ Q
_ b (r _ 0), \:Q _ {ks} (r _ 0) $ are expressions that can be
resulted in by what ever way of regularization of corresponding
integrals. From here it follows that

$$ \frac {({\cal F} _ b) _ {TT}} {({\cal F} _ {ks}) _ {TT}} = \frac {
Q _ b} {Q _ {ks}} \:\frac {d _ 1} {d _ 2} \:\frac {2d _ 1-1} {2d _
2-1} \: t^{2 (d _ 2-d _ 1)}. $$

Assuming $ d _ 2-d _ 1 = O (t) $ at $ T \nearrow T _ {\ast} $ and
eliminating an uncertainty available here, we have

$$ { ({\cal F} _ b) _ {TT}} _ {| T = T _ {\ast}} \ne {({\cal F} _
{ks}) _ {TT}} _ {| T = T _ {\ast}}, $$ and, thus, we show that at
the critical point the existence of a change of phase of the 2-nd
sort is possible (this qualitative conclusion, certainly, does not
depend on a way of regularization). It should be also noticed that
exact values of the critical factors can be determined on
obtaining identities of traces arising in analyzing an appropriate
spectral problem (in [8] the values for half-plane have turned out
to be equal to 1/2.).

{ \bf Conclusion.} In the present paper an exact solution for
Debay-Chukkel's two-dimensio\-\-nal model has been constructed and
its asymptotic form has been found as well. In the nonlinear case
by means of Darboux Transformation method exact solutions have
been calculated and a hypothesis about an possibility of a change
of phase has been put forward and checked. Note that we did not
solve a boundary problem for the equations (4), as such, because
IST, generally speaking, is not adapted for this procedure
{\footnote {Recently some progress in this direction has been
achieved in [19] for a hyperbolic version of $ \sin $ -Gordon
equation.}}. Therefore a conclusion about existence of a change of
phase has, as a whole, a preliminary character. Besides a number
of other important questions arising in the framework of PM such
as question about thermodynamics of investigated system, about
changes of phase corresponding to more complex soliton solutions,
about methods of functionals regularization and others have not
been considered either but will be discussed in future.
\vskip0.3cm {\bf Acknowledgement.}
 The author is extremely grateful to V.D. Lipovsky for
attracting attention to the subject and setting the problem.

\vskip0.5cm

\vskip1cm {\bf REFERENCES} \vskip0.5cm 1.\parbox[t]{12.7cm} {{\em
Frank-Kamenetskii M.D., Anshelevitch V.V. and Lukashin A.V.} Sov.
Phys. Usp., 1987, vol. 151, 4, pp. 595-618.} \vskip0.3cm

2.\parbox[t]{12.7cm} {{\em Joyce G. and Montgomery D.} J. Plasma
Phys., 1973, vol.10, 1, pp. 107-121.} \vskip0.3cm

3.\parbox[t]{12.7cm} {{\em Montgomery D. and Joice G.} Phys.
Fluids, 1974, vol.17, 6, pp. 1139-1145.} \vskip0.3cm

4.\parbox[t]{12.7cm}{{\em Bitsadze A.V.} The equations of
mathematical physics, Moskow, 1982 (in Russian).} \vskip0.3cm

5.\parbox[t]{12.7cm}{{\em Faddeev L.D. and Takhtajan L.A.} The
Hamiltonian Methods in the Theory of Solitons, Berlin, Springer,
1987.} \vskip0.3cm

6.\parbox[t]{12.7cm}{{\em Ablowitz M.J. and Segur H.} Solitons and
Inverse Scattering Transform, Philadelfia, SIAM, 1981.}
\vskip0.3cm

7.\parbox[t]{12.7cm}{{\em Lipovsky V.D. and Nikulichev S.S.}
Vestnik LGU, ser. Fizika, Chimia, 1989, 4, pp. 61-65 (in
Russian).} \vskip0.3cm

8.\parbox[t]{12.7cm}{{\em Gutshabash E.Sh., Lipovsky V.D. and
Nikulichev S.S.} Teoreticheskaja i mathematicheskaia fizika, 1998,
vol.115, 3, pp. 323-348; Preprint nlin.SI/0001012. 2000.}
\vskip0.3cm

9.\parbox[t]{12.7cm}{{\em Gutshabash E.Sh.} Zapiski nauchnych
seminarov POMI, 1998, vol.251, 15, pp. 215-232 (in Russian);
Preprint nlin.SI/0101015. 2001.} \vskip0.3cm

10.\parbox[t]{12.7cm}{{\em Borisov A.B., Kiselev V.V. and Ionov
S.N.} Inverse Scattering Transform for integration sin-Gordon
equation. 1989. Sverdlovsk. VINITI, N0 2015-V89 (in Russian).}
\vskip0.3cm

11.\parbox[t]{12.7cm}{{\em Borisov A.B. and Kiselev V.V.} Inv.
Problem, 1989, vol.5, pp. 959-982.} \vskip0.3cm

12.\parbox[t]{12.7cm}{{\em Borisov A.B.} "Nonlinear exitations and
two-dimensional topological solitons in magnets", Doctorial
dissertation, Inst. Fiz. Metallov, Sverdlovsk, 1986.}} \vskip0.3cm

13.\parbox[t]{12.7cm}{{\em Deserno M., Jimenes-Angeles F., Holm C.
and Lozada-Cassou M.} Preprint cond-mat/0104002. 2001.}
\vskip0.3cm

14.\parbox[t]{12.7cm}{{\em Matveev V.B. and Salle M.A.} Darboux
Transformation and Solitons. Berlin, Springer, 1991.} \vskip0.3cm

15.\parbox[t]{12.7cm}{{\em Landau L.D. and Lifshits E.M.}
Statystical Physics, Moskow, Nauka, 1976 (in Russian).}
\vskip0.3cm

16.\parbox[t]{12.7cm}{{\em Mahan'kov V.G., Rybakov Ju.P. and Sanuk
V.I.}  Sov. Phys. Usp., 1994, vol.164, 2, pp. 121-148.}
\vskip0.3cm

17.\parbox[t]{12.7cm}{{\em Patashinskii A.Z. and Pokrovskii V.L.}
Fluctuating theory of phase transitions, Moskow, Nauka, 1975 (in
Russian).} \vskip0.3cm

18.\parbox[t]{12.7cm}{{\em Shang-keng Ma.}  Modern Theory of
Critical Phenomena, W.a.Benjamin, Inc., 1976.} \vskip0.3cm

19.\parbox[t]{12.7cm}{{\em Leon J. and Spire A.} J. Phys. A: Math.
Gen., 2001, vol. 34, pp. 7359-7380.} \vskip0.4cm

\end{document}